\documentclass[a4paper,11pt]{article}
\usepackage[latin1]{inputenc}
\usepackage{graphicx,color,rotating}
\usepackage{amsmath}
\usepackage{natbib}
\usepackage{hyperref}
\usepackage{float}
\usepackage{multirow}
\usepackage[margin=1in]{geometry}
\usepackage{authblk}
\usepackage{setspace}

\newcommand{\esp}{\mathrm{E}}
\newcommand{\pr}{\mathrm{p}}
\newcommand{\ud}{\mathrm{d}}
\newcommand{\vect}[1]{\boldsymbol{#1}}
\newcommand{\wh}{\widehat}

\newcommand{\vM}{\mathrm{vM}}
\newcommand{\SB}{\mathrm{SB}}
\newcommand{\lin}{\mathrm{lin}}
\newcommand{\MAP}{\mathrm{MAP}}
\newcommand{\MML}{\mathrm{MML}}
\newcommand{\pol}{\kappa,\mu}
\newcommand{\cart}{x,y}
\newcommand{\med}{\mathrm{med}}

\newcommand{\RepFigRes}{Files}

\title{Estimating the concentration parameter of a von Mises distribution: a systematic simulation benchmark}

\author{Guillaume Marrelec\textsuperscript{1,2,*} and Alain Giron\textsuperscript{1,2,*}\\
\footnotesize \textsuperscript{1} Sorbonne Universit{\'e}, CNRS, INSERM, Laboratoire d'imagerie biom{\'e}dicale, LIB, F-75006, Paris, France\\
\footnotesize \textsuperscript{2} Centre de recherches et d'études en sciences des interactions, CR{\'E}SI, Center for Interaction Science, CIS, F-75006, Paris, France\\
\footnotesize \textsuperscript{*} email: firstname.lastname@inserm.fr}

\date{}

\newcommand{\gcite}{\citep}
\newcommand{\gcitet}{\citet}
\newcommand{\gcitep}{\citealp}

\begin{document}

\maketitle

\begin{abstract}
In directional statistics, the von Mises distribution is a key element in the analysis of circular data. While there is a general agreement regarding the estimation of its location parameter $\mu$, several methods have been proposed to estimate the concentration parameter $\kappa$. We here provide a thorough evaluation of the behavior of 12 such estimators for datasets of size $N$ ranging from 2 to 8\,192 generated with a $\kappa$ ranging from 0 to 100. We provide detailed results as well as a global analysis of the results, showing that (1) for a given $\kappa$, most estimators have behaviors that are very similar for large datasets ($N \geq 16$) and more variable for small datasets, and (2) for a given estimator, results are very similar if we consider the mean absolute error for $\kappa \leq 1$ and the mean relative absolute error for $\kappa \geq 1$.
\par
\
\par
\noindent \textit{Keywords:} circular data; von Mises distribution; concentration parameter; estimation; comparison
\end{abstract}

\section{Introduction}

In directional statistics, the von Mises distribution is a key element in the analysis of circular data. Its distribution is given by
\begin{equation} \label{eq:def:vm}
 f ( x ) = \frac{ 1 } { 2 \pi I_0 ( \kappa ) } e ^ { \kappa \cos ( x - \mu ) }, \qquad x \in [0, 2 \pi ),
\end{equation}
where $\mu \in [ 0, 2 \pi )$ is the location parameter, $\kappa \geq 0$ the concentration parameter, and $I_0 ( \kappa )$ the incomplete Bessel integral of the first kind,
\begin{equation}
 I_0 ( \kappa ) = \frac{ 1 } { 2 \pi } \int_0^{2\pi} e ^ { \kappa \cos ( x ) } \, \ud x.
\end{equation}
The von Mises distribution $\vM ( \mu, \kappa )$ is a unimodal distribution with its mode at $x = \mu$. $\kappa = 0$ corresponds to the uniform distribution, and the larger $\kappa$, the more concentrated the distribution is around its mode. The von Mises distribution is a convenient way to model unimodal circular or directional data in many fields of science \gcite{Fisher_NI-1995} and naturally arises in a wide variety of cases \gcite[\S3.5.4]{Mardia-2000}. It can be obtained by conditioning a bivariate normal distribution on the circle, is close to the wrapped normal distribution, and tends to a normal distribution when $\kappa \to \infty$. Also, it is a maximum-entropy distribution and can be characterized by the fact that the maximum likelihood of its location parameter is the sample mean direction. As such, it is an analogue on the circle of the Gaussian distribution. It also appears as the equilibrium distribution of a von Mises process.
\par
While there is a general agreement regarding the estimation of $\mu$ \gcite[\S12.4.1]{Mardia-2000}, estimation of $\kappa$ has generated more research, and several methods have been proposed so far. For a partial review, see, e.g.,  \gcitet[\S12.4.2]{Mardia-2000}. Here, we provide a systematic benchmark of existing estimators. More specifically, we considered 12 estimators: seven frequentist estimators, including the maximum-likelihood estimator \gcite[\S5.3.1]{Mardia-2000}, the marginal maximum-likelihood estimator \gcite{Schou-1978}, two estimators with bias  correction \gcite{Best-1981}, two estimators based on the circular median \gcite{Lenth-1981, Ko-1992}, and an estimator based on the normal approximation \gcite{Abeyasekera-1982}; and five Bayesian estimators \gcite{Dowe-1996}, including three maximum a posteriori (MAP) estimators and two minimum message length (MML) estimators. We assessed the behavior of these 12 estimators for datasets of size $N$ ranging from 2 to 8\,192 generated with a $\kappa$ ranging from 0 to 100. We provide detailed results as well as a more global view of the estimators' performance. Our two main findings are that (1) for a given $\kappa$, most estimators have behaviors that are very similar for large datasets ($N \geq 2^4$) and much more variable for small datasets, and (2) for a given estimator, results are very similar if one consider the mean absolute error for $\kappa \leq 1$ and the mean relative absolute error for $\kappa \geq 1$.
\par
The outline of the manuscript is the following. In Section~\ref{s:rev}, we provide a quick description of the 12 estimators used here. The simulation study itself is detailed in Section~\ref{s:simu}. Further issues are raised in the discussion.

\section{Overview of existing estimators} \label{s:rev}

We here quickly review existing approaches, including frequentist (Section~\ref{ss:rev:freq}) and Bayesian (Section~\ref{ss:rev:bayes}) estimators. All approaches start from a sample $\{ x_1, \dots, x_N \}$ of $N$ independent and identically distributed (i.i.d.) realization of a $\mathrm{vM} ( \mu, \kappa )$ distribution with $\mu$ and $\kappa$ unknown.

\subsection{Frequentist estimators} \label{ss:rev:freq}

Most frequentist estimators of $\kappa$ relie on the likelihood of the parameters. Letting $\bar{R} e ^ { i m }$ be the sample circular mean of the data, \begin{equation} \label{eq:def:moycirc}
 \bar{R} e^{im} = \frac{1}{N} \sum_{ n = 1 } ^ N e ^ { i x_n },
\end{equation}
the likelihood can be expressed from  Equation~(\ref{eq:def:vm}) and the properties of $\bar{R}$ as
\begin{equation} \label{eq:vrais}
 L ( \mu, \kappa ) = \pr ( D | \mu, \kappa ) = \left[ 2 \pi I_0 ( \kappa ) \right] ^ { - N } \exp \left[ \kappa N \bar{R} \cos ( \mu - m ) \right].
\end{equation} 
The maximum-likelihood estimator $\hat{\kappa}$ is the value of $\kappa$ that cancels the derivative of $l ( \mu, \kappa )$. Setting
\begin{equation} \label{eq:def:A}
 A ( \kappa ) = \frac{ I_1 ( \kappa ) } { I_0 ( \kappa ) },
\end{equation}
it is the solution of \gcite[\S5.3.1]{Mardia-2000}
\begin{equation} \label{eq:freq:mv}
 A ( \hat{\kappa} ) = \bar{R}.
\end{equation}
A marginal maximum-likelihood estimate $\tilde{\kappa}$ was also proposed. It is based on the expression of the density of $R = N \bar{R}$ with respect to Lebesgue measure \gcite{Watson_GS-1956, Mardia-1975b},
\begin{equation}
 h_N ( R ; \kappa ) = \frac{ I_0 ( \kappa R ) }{ [ I_0 ( \kappa ) ] ^ N } R h_N ( R; 0 ).
\end{equation}
The maximum $\tilde{\kappa}$ of this expression depends on the relationship between $\bar{R}$ and $N$ \gcite{Schou-1978}
\begin{itemize}
 \item For $0 \leq \bar{R} \leq 1/ \sqrt{N}$, $\tilde{\kappa} = 0$;
 \item For $1 / \sqrt{N} < \bar{R} < 1$, $\tilde{\kappa}$ is the solution of
 \begin{equation} \label{eq:freq:mvm}
  A ( \tilde{\kappa} ) = \bar{R} A ( N \bar{R} \tilde{\kappa} );
 \end{equation}
 \item For $\bar{R} = 1$, there is no maximum (as it would correspond to $\tilde{\kappa} = \infty$).
\end{itemize}
Using both a calculation based on a Taylor expansion around $\esp ( \bar{R} )$ of $\hat{\kappa} = A ^ { - 1 } ( \bar{R} )$ from  Equation~(\ref{eq:freq:mv}) and a simulation study, \gcitet{Best-1981} showed that the (regular) maximum-likelihood estimator $\hat{\kappa}$  of Equation~(\ref{eq:freq:mv}) can be strongly biased for small $\kappa$ and $N$. They proposed to correct for this bias using an approximate expansion, leading to
\begin{equation} \label{eq:def:BF1}
 \hat{\kappa}_1^* = \left\{ \begin{array}{cl}
 \max \left\{ \hat{\kappa} - \frac{ 2 }{ N \hat{ \kappa } }, 0 \right\} & \mbox{for $\hat{\kappa} < 2$} \\
 \frac{ ( N - 1 ) ^ 3 } { N ^ 3 + N } \hat{\kappa} & \mbox{for $\hat{\kappa} \geq 2$.}
\end{array} \right.
\end{equation}
They also proposed a jackknife correction of the bias,
\begin{equation} \label{eq:def:BF2}
 \hat{\kappa}_2^* = \max \left\{ N \hat{\kappa} - \frac{N-1}{N} \sum_{ n = 1 } ^ N \hat{\kappa}_{ - n }, 0 \right\}.
\end{equation}
By analogy with the median absolute deviation, \gcitet{Lenth-1981} proposed an estimator based on the median
\begin{equation} \label{eq:def:median1}
 \hat{\kappa}_L = \frac{ 0.6724 } { \mathrm{median}_n \left\{ 2 \left[ 1 - \cos ( x_n - \hat{\mu}_{\med}) \right] \right\} },
\end{equation}
where $\hat{\mu}_{\med}$ is the circular median \gcite[\S2.2.2]{Mardia-2000}. \gcitet{Ko-1992} improved this approach by replacing the expectation and average of Equation~(\ref{eq:freq:mv}) with median and sampling median \gcite[see also][]{Ducharme-1990}, yielding an estimator $\hat{\kappa}_{\med}$ such that
\begin{equation} \label{eq:def:median2}
 \int_{-1}^{\mathrm{median}_n [ \cos ( x_n - \hat{\mu}_{\SB} ) ]} \frac{1}{\pi I_0 ( \hat{\kappa}_{\med} )} \frac{ e^{\hat{\kappa}_{\med} t} } { \sqrt{ 1 - t^2 } } = \frac{1}{2},
\end{equation}
where $\hat{\mu}_{\SB}$ is any estimator of $\mu$ that is standardized bias (SB) robust, such as the circular median or the least median square (LMS) estimator. 
\par
Finally, \gcitet{Abeyasekera-1982} made use of the fact that a $\mathrm{vM} ( \mu, \kappa )$ distribution is well approximated by the normal distribution $\mathcal{N} ( \mu, \kappa ^ { - 1 } )$ for large $\kappa$ and therefore proposed as estimator of $\kappa$ the unbiased estimator for $\sigma^{-2}$, leading to the so-called linear estimator for $N > 3$
\begin{equation} \label{eq:def:linear}
 \hat{\kappa}_{\lin} = \left[ \frac{1}{N-3} \sum_{ n = 1 } ^ N ( x_n - \bar{x} ) ^ 2 \right] ^ { - 1 },
\end{equation}
where $\bar{x}$ is the linear mean, that is, the usual arithmetic mean
$$\bar{x} = \frac{1}{N} \sum_{ n = 1 } ^ N x_n.$$
Note that there is an issue induced by the discontinuity at 0, which is dealt with by using a technical trick. They also proposed an improved estimator $\hat{\kappa}_{\lin}^*$ where the bias is corrected by jackknifing the linear estimator.

\subsection{Bayesian estimators} \label{ss:rev:bayes}

In a Bayesian setting, all the information that can be inferred about the parameters $\mu$ and $\kappa$ from a dataset $D$ is summarized in the posterior distribution $\pr (\mu, \kappa | D )$, which, according to Bayes' theorem, can be expressed as
\begin{equation}
 \pr ( \mu, \kappa | D ) \propto \pr ( \mu, \kappa ) \, \pr ( D | \mu, \kappa ),
\end{equation}
where ``$\propto$'' relates two expressions that are proportional. $\pr ( D | \mu, \kappa )$ is the data likelihood, already expressed in Equation~(\ref{eq:vrais}). $\pr ( \mu, \kappa )$ is the prior distribution, which translates the information we have about the parameters before we have the data. A common approach is to assume no prior dependence between parameters, so that $\pr ( \mu, \kappa ) = \pr ( \mu ) \, \pr ( \kappa )$.
$\pr ( \mu )$ is classically set as a noninformative uniform distribution on the circle, $\pr ( \mu ) = 1 / 2 \pi$.
As for $\pr ( \kappa )$, \gcitet{Dowe-1996} considered the following two prior distributions
\begin{equation} \label{eq:prior}
 h_2 ( \kappa ) = \frac{2}{\pi(1+\kappa^2)} \qquad \mbox{and} \qquad h_3 ( \kappa ) = \frac{\kappa}{(1+\kappa^2)^{\frac{3}{2}}}, \qquad \kappa \geq 0.
\end{equation}
The posterior distribution $\pr ( \mu, \kappa | D )$ then yields
\begin{equation}
 \pr ( \mu, \kappa | D ) \propto h_i ( \kappa ) \, \left[ 2 \pi I_0 ( \kappa ) \right] ^ { - N } \exp \left[ \kappa N \bar{R} \cos ( \mu - m ) \right].
\end{equation}
They then proposed to use the maximum a posteriori (MAP) estimators
\begin{equation} \label{eq:def:MAP}
 \kappa_{\MAP,i}^{\pol} = \mathrm{arg}_{ \kappa } \max _{ \mu, \kappa } \pr ( \mu, \kappa | D ) = \mathrm{arg}_{ \kappa \geq 0 } \max \left\{ \ln h_i ( \kappa ) - N \ln [ 2 \pi I_0 ( \kappa ) ] + N \bar{R} \kappa \right\}, \qquad i = 2, 3.
\end{equation}
For $h_3 ( \kappa )$, they also proposed to consider estimating $\kappa$ as the MAP of the distribution of $[ x = \kappa \cos ( \mu ), y = \kappa \sin ( \mu ) ]$, leading to
\begin{equation} \label{eq:def:MAPcart}
 \kappa_{\MAP,3}^{\cart} = \max_{ \kappa \geq 0 } \left\{ \ln \left[ \frac{ h_3 ( \kappa ) } { \kappa } \right] - \ln [ 2 \pi I_0 ( \kappa ) ] + N \bar{R} \kappa \right\}.
\end{equation}
Finally, they proposed two minimum message length (MML) estimators. Minimizing the message length is equivalent to maximizing
\begin{equation}
 \frac{ \pr ( \mu, \kappa | D ) }{ \sqrt{ \det F ( \mu, \kappa ) } },
\end{equation}
where $\det F ( \mu, \kappa )$ is the determinant of the Fisher matrix for the von Mises distribution. The resulting estimators for $h_2 ( \kappa )$ and $h_2 ( \kappa )$ are
\begin{equation} \label{eq:def:MML2}
 \kappa_{\MML,2} = \arg \max_{ \kappa \geq 0 } \frac{ h_2 ( \kappa ) \, \pr ( \vect{x} | \mu, \kappa ) } { \sqrt{ \left[ \kappa A ( \kappa ) + \frac{3}{N \pi ^ 2} \right] A' ( \kappa ) } }
\end{equation}
and
\begin{equation} \label{eq:def:MML3}
 \kappa_{\MML,3} = \arg \max_{ \kappa \geq 0 } \frac{ h_3 ( \kappa ) \, \pr ( \vect{x} | \mu, \kappa ) } { N \sqrt{ \kappa A ( \kappa ) A' ( \kappa ) } },
\end{equation}
respectively.

\subsection{Existing simulation studies}

We found three existing simulations studies assessing the relative performance of estimators of the concentration parameter: \gcitet{Best-1981}, \gcitet{Abeyasekera-1982}, and \gcitet{Dowe-1996}. Their scope and methodology are summarized in Table~\ref{tab:sim:bib}. \gcitet{Best-1981} found that $\hat{\kappa}_1^*$ and $\hat{\kappa}_2^*$ were similar and better than $\hat{\kappa}$ for small values of $\kappa$ and $N$. $\hat{\kappa}_1^*$ was less biased than $\hat{\kappa}$ and $\hat{\kappa}_2^*$ was less median biased. However, the bias remained for small $\kappa$ and $N$. \gcitet{Abeyasekera-1982} found that $\hat{\kappa}_2^*$ was better than $\hat{\kappa}$, $\tilde{\kappa}$ and $\tilde{\kappa}^*$; that $\tilde{\kappa}$ was better than $\hat{\kappa}$ for small values of $N$ and worse when $N$ was large; and, finally, that $\tilde{\kappa}$ and $\tilde{\kappa}^*$ had similar behaviors. Finally, \citet{Dowe-1996} found that Bayesian methods outperformed frequentist methods, with a limited effect of the prior distribution on the behavior.

\begin{sidewaystable}[!htbp]
 \small
 \caption{\textbf{Existing simulation.} Summary of scope and methodology. $N$ is the sample size and $M$ the number of samples used to compute mean errors. MAE: mean absolute error; MSE: mean squared error; MKL: mean Kullback--Leibler divergence; MRAE: mean relative absolute error.}
 \begin{tabular}{cccccccc}
  Article & $\kappa$ & $N$ & $M$ & Estimators & Measures of fit \\
  \hline
  \gcitet{Best-1981} & $\{ 0.1, 0.5, 1, 2.5, 5 \}$ & $\{ 10, 20, 100 \}$ & $\geq 1\,000$ & $\hat{\kappa}$ & qualitative on sampling features \\
  & $\{ 0.1, 0.5, 1, 2.5, 5, 10, 100, 500 \}$ & $\{ 10, 20 \}$ & $\geq 1\,000$ & $\hat{\kappa}_1^*$, $\hat{\kappa}_2^*$ & qualitative on sampling features \\
  & $\{ 0.1, 0.5, 1, 2.5, 5 \}$ & 100 & $\geq 1\,000$ & $\hat{\kappa}_1^*$, $\hat{\kappa}_2^*$ & qualitative on sampling features \\
  & $\{ 0.01, 0.2, 0.3, 0.4, 0.75 \}$ &  $\{ 10, 20 \}$ & $\geq 1\,000$ & $\hat{\kappa}_1^*$, $\hat{\kappa}_2^*$ & qualitative on sampling features \\
  & $\{ 0.01, 0.2, 0.3 \}$ &  100 & $\geq 1\,000$ & $\hat{\kappa}$, $\hat{\kappa}_1^*$, $\hat{\kappa}_2^*$ & qualitative on sampling features \\
  \hline
  \gcitet{Abeyasekera-1982} & $\{ 0.1, 0.5, 1, 2, 3, 4, 5, 7.5 \}$ & $\{ 10, 20, 30, 100 \}$ & 1\,000 & $\hat{\kappa}$, $\tilde{\kappa}$, $\hat{\kappa}_2^*$, $\tilde{\kappa}^*$ & sample bias and MSE \\  
  \hline
  \gcitet{Dowe-1996} & $\{ 0, 0.5, 1, 10 \}$ & $\{ 2, 5 \}$ & 1\,000 & $\hat{\kappa}$, $\tilde{\kappa}$, $\kappa_{\MAP,2}$, $\kappa_{\MAP,3}$ & MAE, MSE, MKL \\
  & $\{ 0, 1, 10 \}$ & $\{ 25, 100 \}$ & 1\,000 & $\hat{\kappa}$, $\tilde{\kappa}$, $\kappa_{\MAP,2}$, $\kappa_{\MAP,3}$ & MAE, MSE, MKL \\
  \hline
  our simulation & $\{ 0, 0.01, 0.1, 1, 10, 100 \}$ & $\{ 2, 4, 8, 16, \dots, 8\,192 \}$ & 1\,000 & see Table~\ref{tab:simu:estim} & MAE, MRAE
 \end{tabular} \label{tab:sim:bib}
\end{sidewaystable}

\section{Simulation study} \label{s:simu}

\subsection{Data}

We generated samples from von Mises distributions. More specifically, for each of the $Q = 6$ values of $\kappa \in \{ 0, 10^{-2}, 10^{-1}, 1, 10, 10^2 \}$, we generated $M = 1\,000$ ``maximal'' datasets. Each dataset was composed of $N_{\max}$ i.i.d. realizations of a $\vM ( \mu, \kappa )$ distribution with $\mu$ uniformly distributed on the circle and $N_{\max} = 2 ^ L$ with $L = 13$ (i.e., $N_{\max} = 8\,192$). Each sample was generated using \gcitet{Berens-2009}'s \texttt{CircStat}. This procedure therefore generated a total of $Q \times M = 6\,000$ maximal datasets of size $8\,192$. From each maximal dataset, we then extracted $L$ datasets composed of the $2 ^ l$ first data points with $l \in \{ 1, \dots, L \}$. This gave us a total of $Q \times M \times L = 78\,000$ datasets on which inference was performed. On each dataset, estimation of $\kappa$ was performed by application of each of the $J = 12$ estimators mentioned above (see also Table~\ref{tab:simu:estim} for a summary). We therefore ended with  a set of $Q \times M \times L \times J = 936\,000$ estimates.

\begin{table}[!htbp]
 \centering
 \begin{tabular}{ccc}
  \textbf{Estimator} & \textbf{Equation} & \textbf{Identification} \\
  \hline
  $\hat{\kappa}$ & (\ref{eq:freq:mv}) & \texttt{jML} \\
  $\tilde{\kappa}$ & (\ref{eq:freq:mvm} & \texttt{mML} \\
  $\hat{\kappa}_1^*$ & (\ref{eq:def:BF1}) & \texttt{BF1} \\
  $\hat{\kappa}_2^*$ & (\ref{eq:def:BF2}) & \texttt{BF2} \\
  $\hat{\kappa}_L$ & (\ref{eq:def:median1}) & \texttt{median-1} \\
  $\hat{\kappa}_{\med}$ & (\ref{eq:def:median2}) & \texttt{median-2} \\
  $\hat{\kappa}_{\lin}$ & (\ref{eq:def:linear}) & \texttt{linear} \\
  $\hat{\kappa}_{\MAP,2}^{\pol}$ & (\ref{eq:def:MAP}) & \texttt{BayesEst-2-jMAP-km} \\
  $\hat{\kappa}_{\MAP,3}^{\pol}$ & (\ref{eq:def:MAP}) & \texttt{BayesEst-3-jMAP-km} \\
  $\hat{\kappa}_{\MAP,3}^{\cart}$ & (\ref{eq:def:MAPcart}) & \texttt{BayesEst-3-jMAP-xy} \\
  $\hat{\kappa}_{\MML,2}$ & (\ref{eq:def:MML2}) & \texttt{MML-2} \\
  $\hat{\kappa}_{\MML,3}$ & (\ref{eq:def:MML3}) & \texttt{MML-3}
 \end{tabular}
 \caption{\textbf{Simulation study.} Summary of estimators used.} \label{tab:simu:estim}
\end{table}

\subsection{Evaluation}

To assess the behavior of the various estimations, we proceeded as follows. For estimation $\kappa_{jlmq}$, obtained by application of estimator $j$ to the $2^l$ first data points of simulation $m$ associated with the $q$th value of $\kappa$, we first considered the absolute error of the estimation compared to the true value
\begin{equation} \label{eq:sim:errabs}
 \epsilon_{jlmq} = | \hat{\kappa}_{jlmq} - \kappa_q |.
\end{equation}
We then summarized the results over the $M$ simulations by computing the mean absolute error (MAE), defined as a sampling average
\begin{equation} \label{eq:sim:mae}
 \overline{\epsilon}_{jlq} = \frac{1}{M} \sum_{ m = 1 } ^ M \epsilon_{jlmq}.
\end{equation}
In the same fashion, we also considered the relative error, where the above error of Equation~(\ref{eq:sim:errabs}) is normalized by the true value 
\begin{equation}
 \epsilon_{jlmq}' = \frac{ | \hat{\kappa}_{jlmq} - \kappa_q | } { \kappa_q },
\end{equation}
and the mean relative absolute error (MRAE) as its sampling average over the $M$ simulations
\begin{equation} \label{eq:sim:mrae}
 \overline{\epsilon'}_{jlq} = \frac{1}{M} \sum_{ m = 1 } ^ M \epsilon_{jlmq}'.
\end{equation}

\subsection{Results}

Detailed results can be found in the supplemental material, in the form of summary statistics (supplemental material, Tables~1--12) and graphs (supplemental material, Figures~1--6). We here focus on some key features of the behavior: computation time, algorithmic failures, the global trend as a function of $N$, the behavior for large samples ($N \geq 2^4$), and the small-sample behavior.

\paragraph{Computation time.}

\begin{figure}[!htbp]
 \centering
 \includegraphics[width=0.7\linewidth]{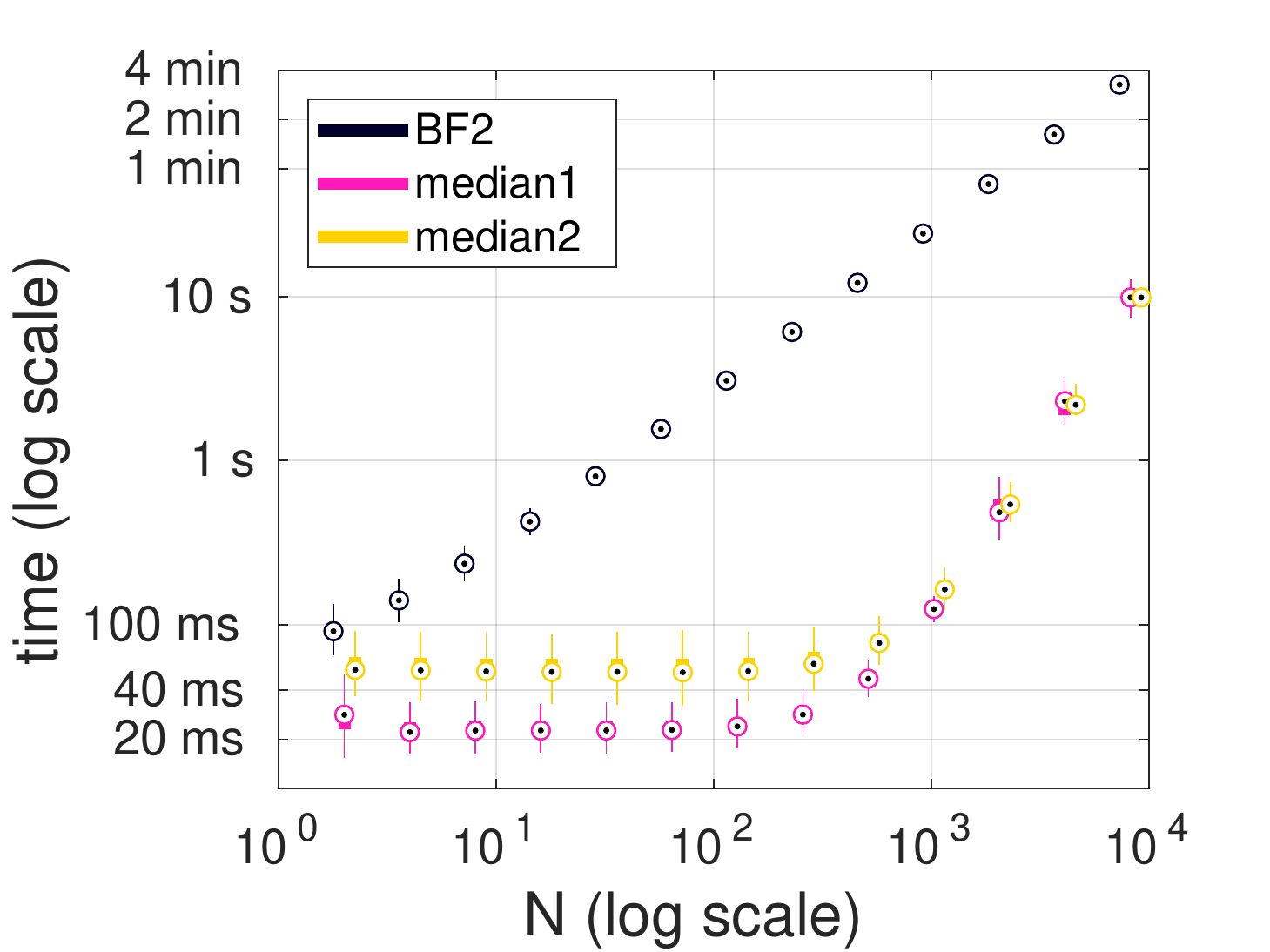}
 \caption{\textbf{Simulation study.} Computation time for estimators that were influenced by sample size. Boxplots (median and $[25\%, 75\%]$ percentile interval) across values of $\kappa$ and simulations.} \label{fig:sim:temps2}
\end{figure}

The detailed computation times can be found in the supplemental material, Tables 1--12, columns 3 and 4. For all estimators, the actual value of $\kappa$ seemed to have very limited influence on the computation time. Similarly, for most estimators (\texttt{jML}, \texttt{mML}, \texttt{BF1}, \texttt{linear}, \texttt{MML-3}, \texttt{BayesEst-2-jMAP-km}, \texttt{BayesEst-3-jMAP-km}, and \texttt{BayesEst-3-jMAP-xy}), the computation time was not a function of the sampling size; it was of the order of 40~ms for \texttt{BF1} and of 20~ms for the others (see Table~\ref{tab:sim:temps}). Three estimators had computation times that were a function of data size: \texttt{BF2}, \texttt{median1}, and \texttt{median2} (see Figure~\ref{fig:sim:temps2}). For \texttt{BF2}, the computation time was roughly linear, with a doubling of the data size corresponding to a doubling of the computation time. For \texttt{median1} and \texttt{median2}, the computation time was roughly constant up to $N = 2 ^7$ and then increased exponentially.

\begin{table}[!htbp]
 \centering
 \caption{\textbf{Simulation study.} Computation time for estimators that were not influenced by sample size. Mean $\pm$ standard deviation across values of $\kappa$, $N$, and simulations.} \label{tab:sim:temps}
 \begin{tabular}{cc}
  \textbf{Estimator} & \textbf{Time (mean $\pm$ std dev)} \\
  \hline
  \texttt{jML} & $23.9 \pm 5.6$ ms \\
  \texttt{mML}l & $23.9 \pm 5.6$ ms \\
  \texttt{BF1} & $46.9 \pm 8.5$ ms \\
  \texttt{linear} & $24.0 \pm 5.6$ ms \\
  \texttt{MML-2} & $24.4 \pm 5.7$ ms \\
  \texttt{MML-3} & $24.5 \pm 5.7$ ms \\
  \texttt{BayesEst-2-jMAP-km} & $24.0 \pm 5.5$ ms \\
  \texttt{BayesEst-3-jMAP-km} & $24.0 \pm 5.4$ ms \\
  \texttt{BayesEst-3-jMAP-xy} & $24.0 \pm 5.3$ ms \\
 \end{tabular}

\end{table}

\paragraph{Failures.}

We first report cases where methods failed to compute an estimate (see Table~\ref{tab:sim:pb}). This happened for \texttt{median-2} (corresponding to $\hat{\kappa}_L$) and \texttt{linear} (corresponding to $\hat{\kappa}_{\lin}$). By definition, $\hat{\kappa}_{\lin}$ of Equation~(\ref{eq:def:linear}) is not defined for $N = 2$. Also, the second median estimator, $\hat{\kappa}_{\med}$, was not always defined, as they were cases for which the integral of Equation~(\ref{eq:def:median2}) was smaller than $1/2$ for all $\kappa \geq 0$ (and, in particular for $\kappa = 0$). Cases corresponding to failures were removed to compute the summary statistics $\overline{\epsilon}$ and $\overline{\epsilon'}$.

\begin{table}[!htbp]
 \centering
 \caption{\textbf{Simulation study.} Summary of number of failures encountered for all $\tau$ and $N$. No failure was encountered for \texttt{jML}, \texttt{mML}, \texttt{BF1}, \texttt{BF2}, \texttt{median1}, \texttt{MML-2}, \texttt{MML-3}, \texttt{BayesEst-2-km}, \texttt{BayesEst-3-km}, and \texttt{BayesEst-3-xy}.}
 \label{tab:sim:pb}
 \begin{tabular}[b]{ccc}
  \ & & \\
  \begin{tabular}[t]{cccc}
   $\tau$ & $N$ & \texttt{median2} & \texttt{linear} \\
   \hline
   0 & $2$ & $0$ & $1000$ \\
   & $4$ & $73$ & $0$ \\
   & $8$ & $109$ & $0$ \\
   & $16$ & $165$ & $0$ \\
   & $32$ & $175$ & $0$ \\
   & $64$ & $194$ & $0$ \\
   & $128$ & $201$ & $0$ \\
   & $256$ & $200$ & $0$ \\
   & $512$ & $211$ & $0$ \\
   & $1024$ & $214$ & $0$ \\
   & $2048$ & $193$ & $0$ \\
   & $4096$ & $241$ & $0$ \\
   & $8192$ & $229$ & $0$ \\
   \hline
   0.01 & $2$ & $0$ & $1000$ \\
   & $4$ & $71$ & $0$ \\
   & $8$ & $105$ & $0$ \\
   & $16$ & $150$ & $0$ \\
   & $32$ & $157$ & $0$ \\
   & $64$ & $189$ & $0$ \\
   & $128$ & $185$ & $0$ \\
   & $256$ & $193$ & $0$ \\
   & $512$ & $194$ & $0$ \\
   & $1024$ & $211$ & $0$ \\
   & $2048$ & $216$ & $0$ \\
   & $4096$ & $232$ & $0$ \\
   & $8192$ & $249$ & $0$
 \end{tabular}
 & \ & \begin{tabular}[t]{cccc}
  $\tau$ & $N$ & \texttt{median2} & \texttt{linear} \\
  \hline
  0.1 & $2$ & $0$ & $1000$ \\
  & $4$ & $51$ & $0$ \\
  & $8$ & $114$ & $0$ \\
  & $16$ & $142$ & $0$ \\
  & $32$ & $138$ & $0$ \\
  & $64$ & $179$ & $0$ \\
  & $128$ & $187$ & $0$ \\
  & $256$ & $187$ & $0$ \\
  & $512$ & $203$ & $0$ \\
  & $1024$ & $193$ & $0$ \\
  & $2048$ & $205$ & $0$ \\
  & $4096$ & $210$ & $0$ \\
  & $8192$ & $231$ & $0$ \\
  \hline
  1 & $2$ & $0$ & $1000$ \\
  & $4$ & $29$ & $0$ \\
  & $8$ & $34$ & $0$ \\
  & $16$ & $54$ & $0$ \\
  & $32$ & $46$ & $0$ \\
  & $64$ & $42$ & $0$ \\
  & $128$ & $52$ & $0$ \\
  & $256$ & $57$ & $0$ \\
  & $512$ & $61$ & $0$ \\
  & $1024$ & $67$ & $0$ \\
  & $2048$ & $58$ & $0$ \\
  & $4096$ & $53$ & $0$ \\
  & $8192$ & $57$ & $0$ \\
  \hline
  10 & $2$ & $0$ & $1000$ \\
  \hline
  100 & $2$ & $0$ & $1000$ \\
 \end{tabular}
\end{tabular}
\end{table}

\paragraph{Evolution as a function of $N$.}

\begin{sidewaysfigure}[!htbp]
 \centering
 \begin{tabular}{ccc}
  $\kappa = 0$ & $\kappa = 0.01$ & $\kappa = 0.1$ \\
  \includegraphics[width=0.3\linewidth]{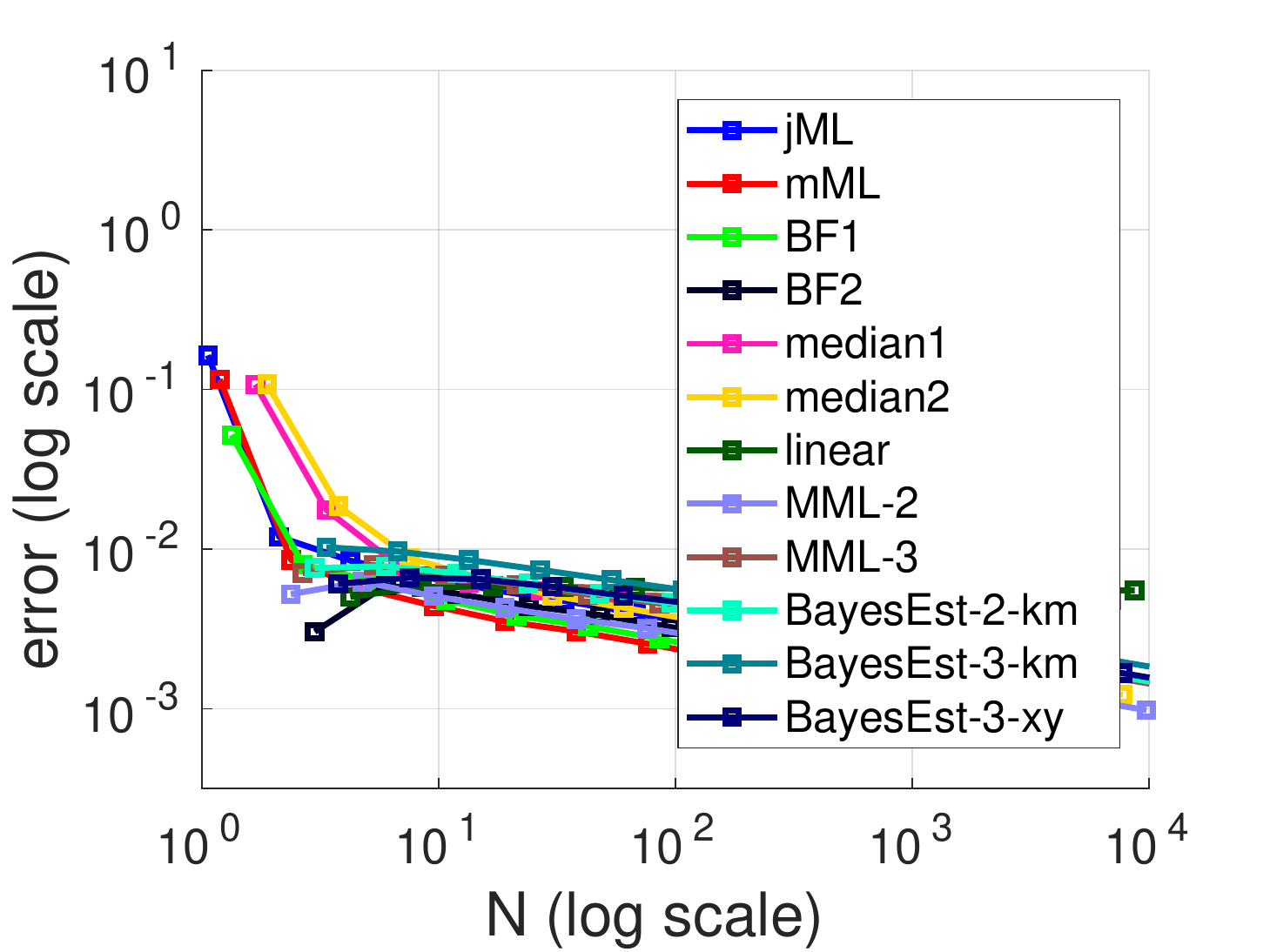}
  & \includegraphics[width=0.3\linewidth]{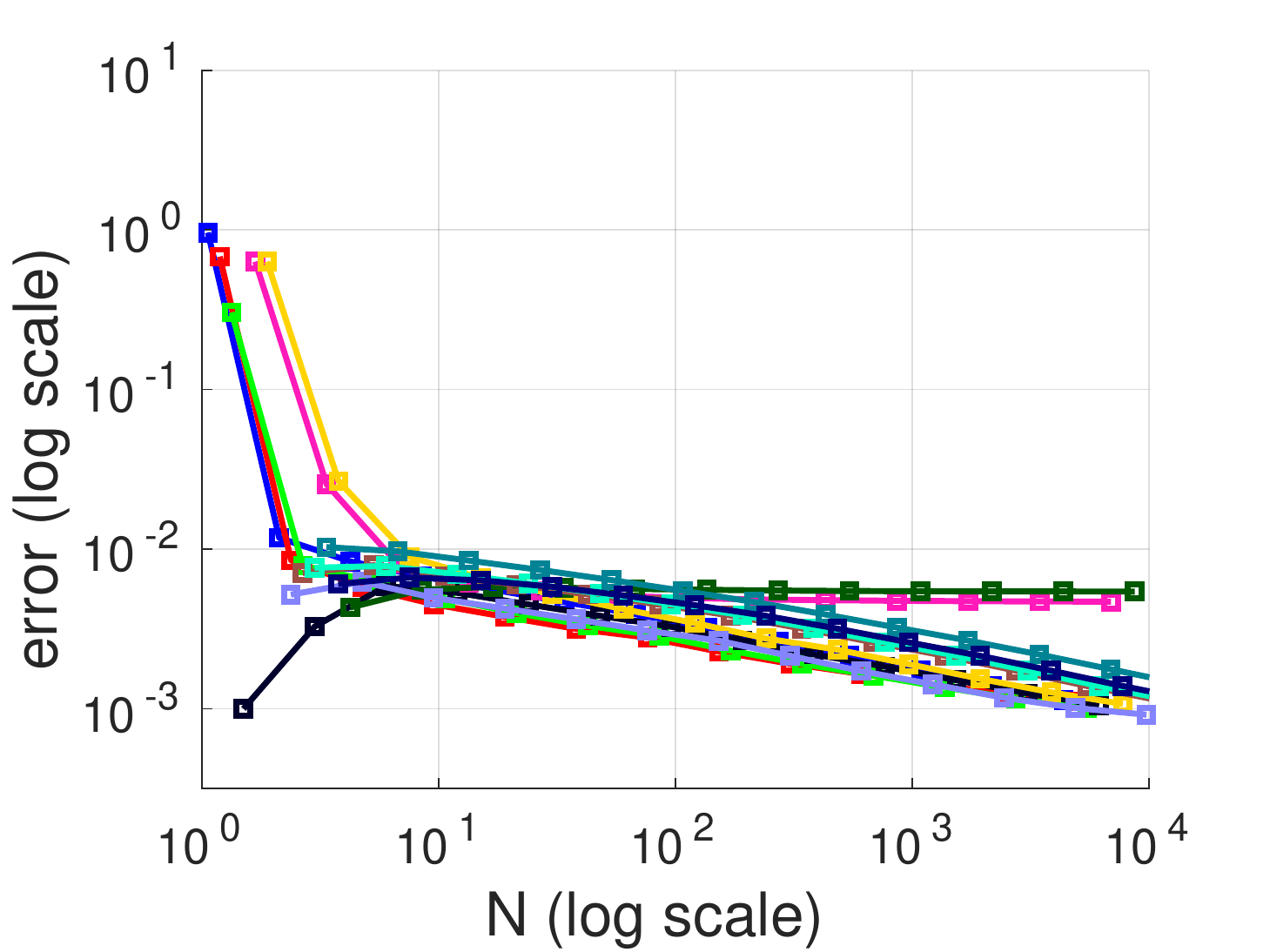}
  & \includegraphics[width=0.3\linewidth]{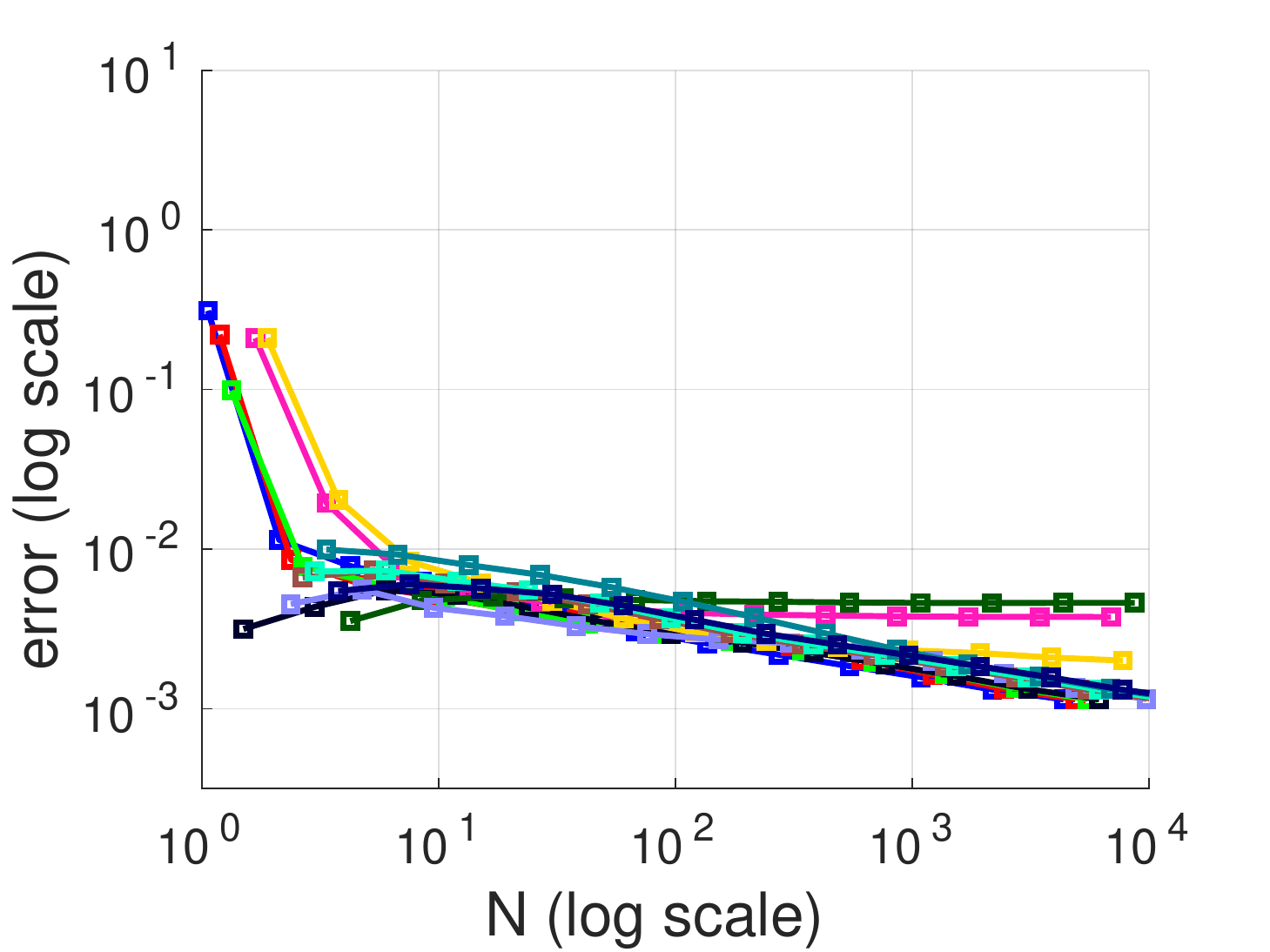} \\
  \\
  \\
  $\kappa = 1$ & $\kappa = 10$ & $\kappa = 100$ \\
  \includegraphics[width=0.3\linewidth]{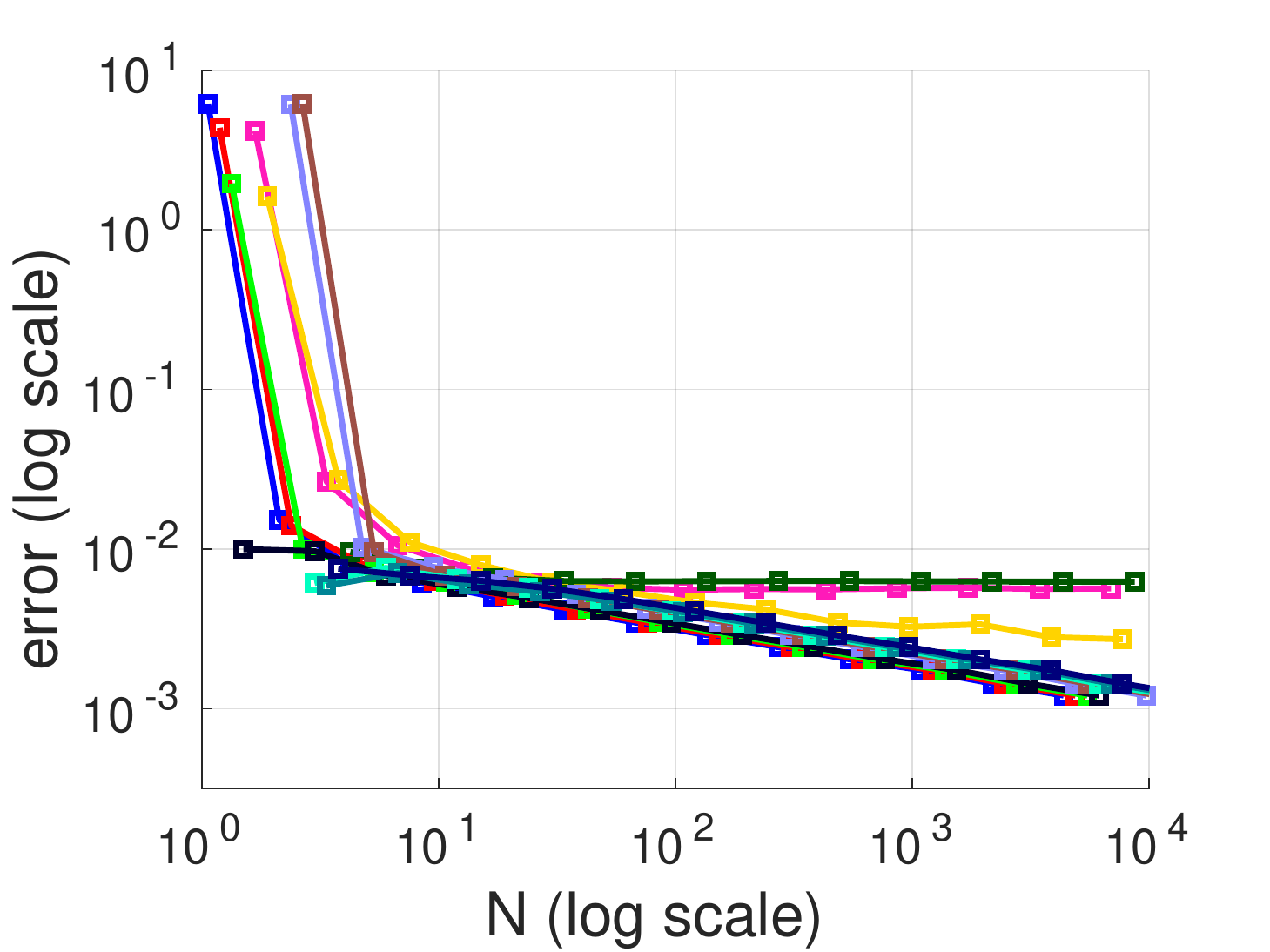}
  & \includegraphics[width=0.3\linewidth]{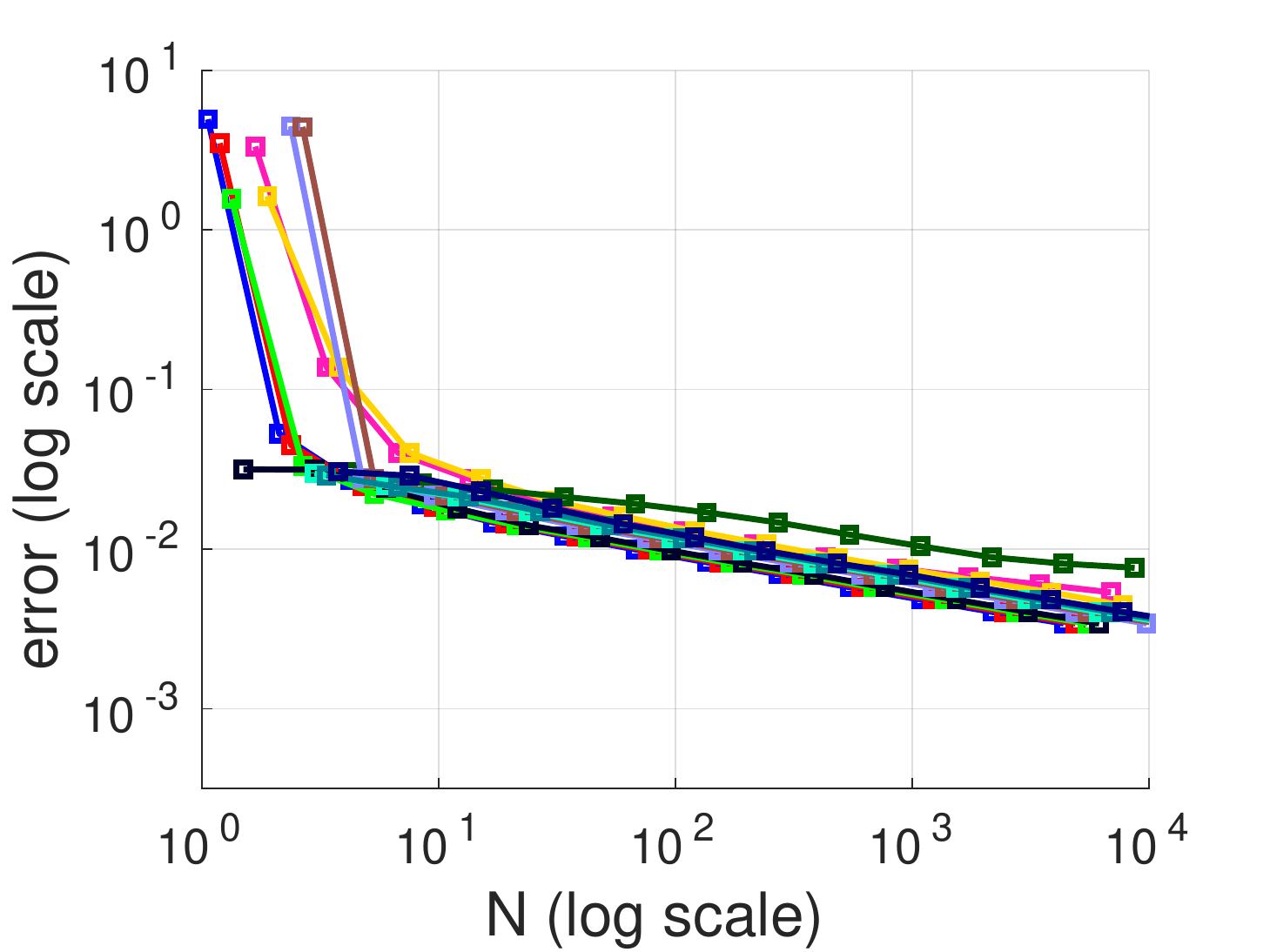}
  & \includegraphics[width=0.3\linewidth]{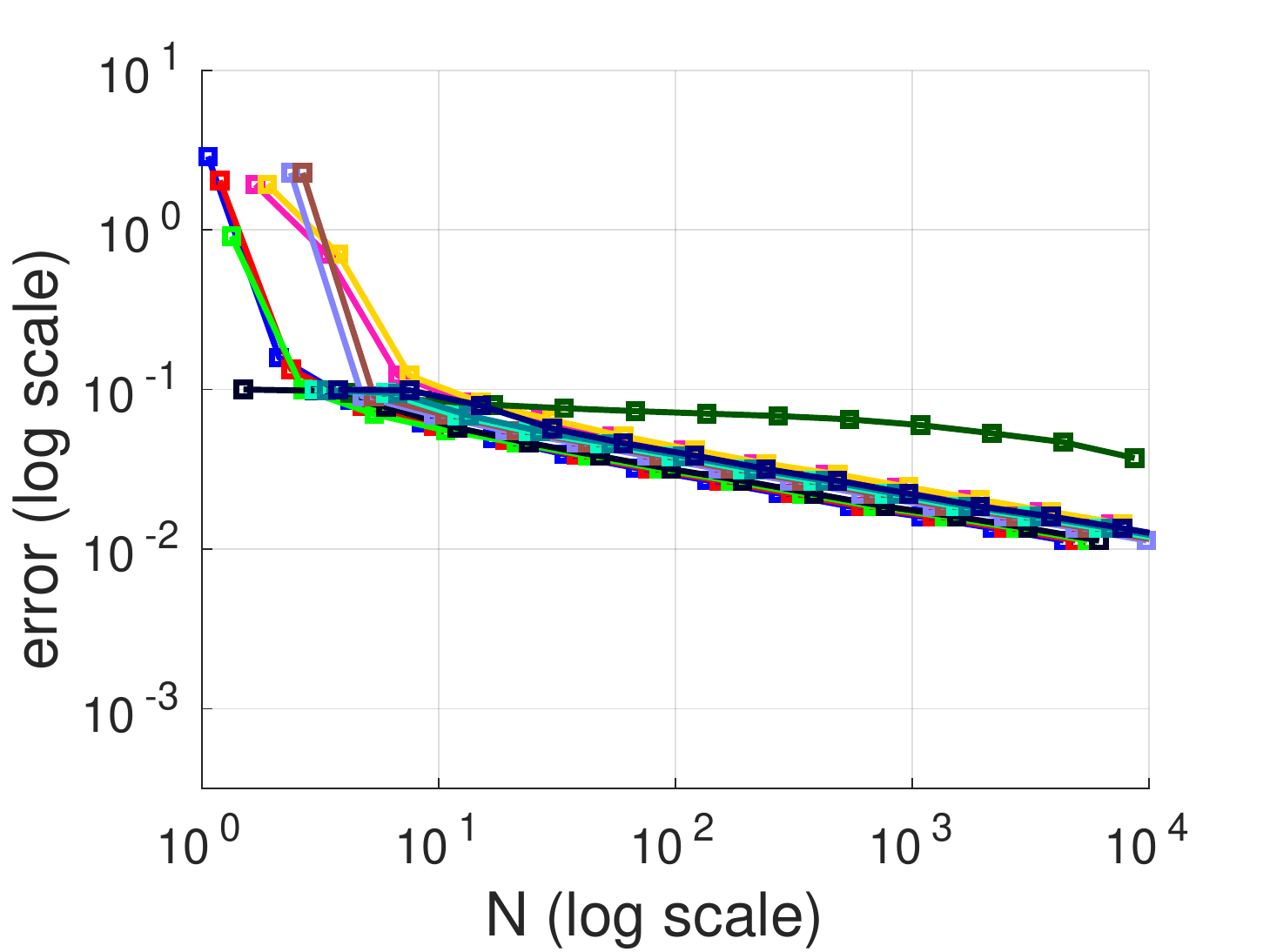} \\
 \end{tabular}
 \caption{\textbf{Simulation study.} Average error $\overline{\epsilon}$ for all values of $\kappa$ and all estimators.} \label{fig:sim:kappa}
\end{sidewaysfigure}

Graphs corresponding to $\overline{\epsilon}$ and $\overline{\epsilon'}$ for all estimators gathered by value of $\kappa$ can be found in Figure~\ref{fig:sim:kappa}. Globally, most estimators improved as the sample size increased. One clear exception was \texttt{BF2} for small $N$ and low $\tau$. In that particular case, the method tended to estimate $\tau$ as equal to 0, leading to an error of the order of $\tau$, with a correction toward a more normal trend as the data size increased. This behavior was also observed to a much lesser extend for the maximum message length estimators (\texttt{MML-2} and \texttt{MML-3}) and the Bayesian estimators (\texttt{BayesEst-2-jMAP-km}, \texttt{BayesEst-3-jMAP-km}, and \texttt{BayesEst-3-jMAP-xy}).

\paragraph{Large-sample behavior.}

We observed quite homogeneous results for large datasets ($N \geq 2^4$), both within- and between estimators. First, the mean absolute error roughly decreased linearly in log-log coordinates for most estimators. The two exceptions were  \texttt{median1} for low $\tau$, where the estimator seemed to reach a plateau independent of $N$, and \texttt{linear}, for which data samples of increasing size did not seem to improve estimation. Also, the performance of a given method for different values of $\kappa$ were quite similar when one considered the mean absolute error $\overline{\epsilon}$ for $\kappa \leq 1$ and the mean relative absolute error $\overline{\epsilon'}$ for $\kappa \geq 1$. The main exception was \texttt{median1} due to the difference in behavior for low values versus high values of $\tau$ mentioned earlier in the paragraph. For the other estimators, the slope of $\log_{10} ( \overline{\epsilon} )$ as a function of $\log_{10} ( N )$ was found to be around $-1/2$, corresponding to a decrease of mean (relative) absolute error in $1/\sqrt{N}$. Finally, most estimators behaved quite similarly in this large-sample regime, at the exception of \texttt{median1}, \texttt{linear} and, to a lesser extend, \texttt{median2}.
\par
\begin{figure}[!htbp]
 \centering
 \begin{tabular}{cc}
  (a) estimated slope $\alpha$ & (b) estimated intercept $\beta$ \\
  \includegraphics[width=0.5\linewidth]{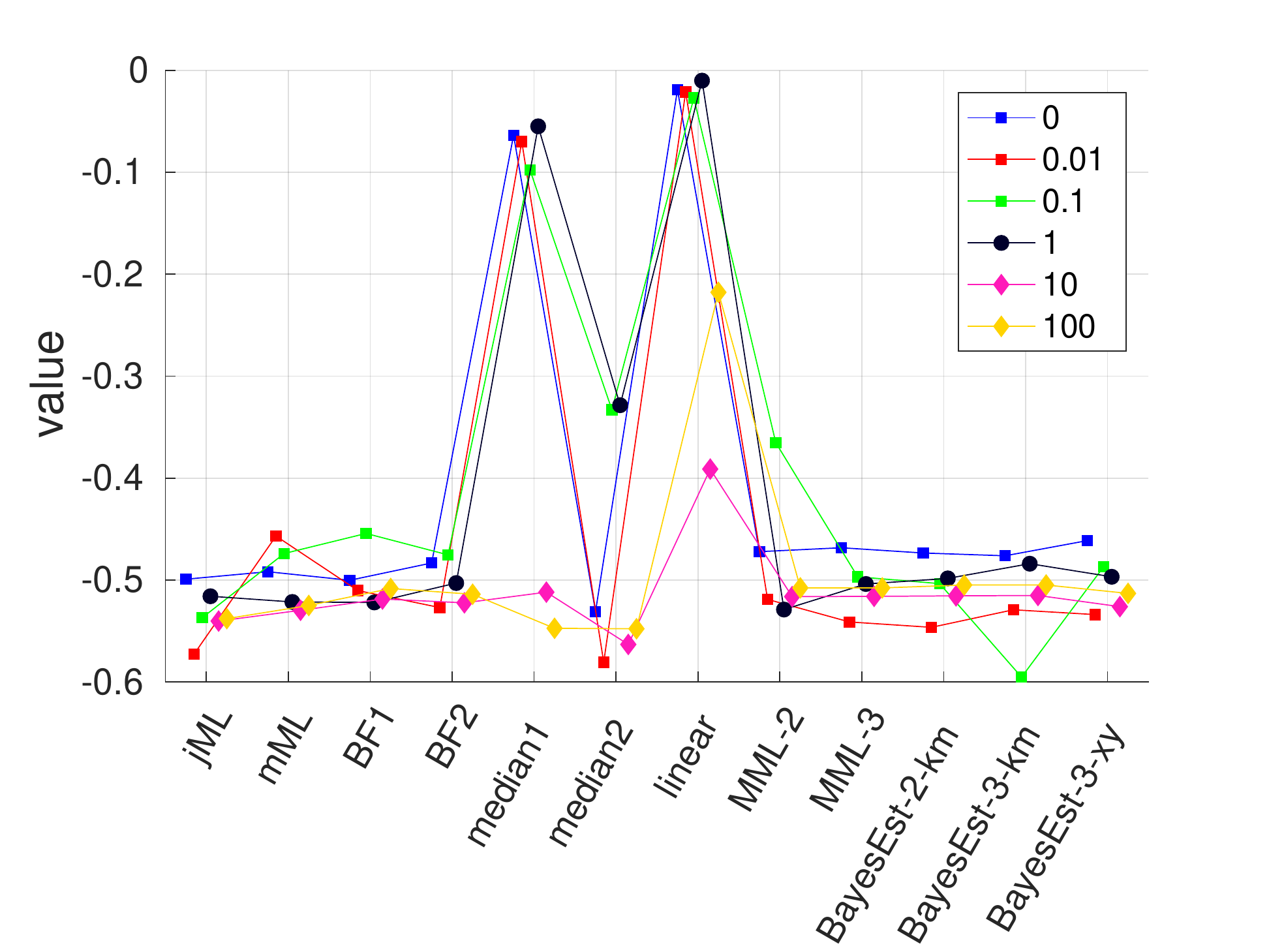}
  & \includegraphics[width=0.5\linewidth]{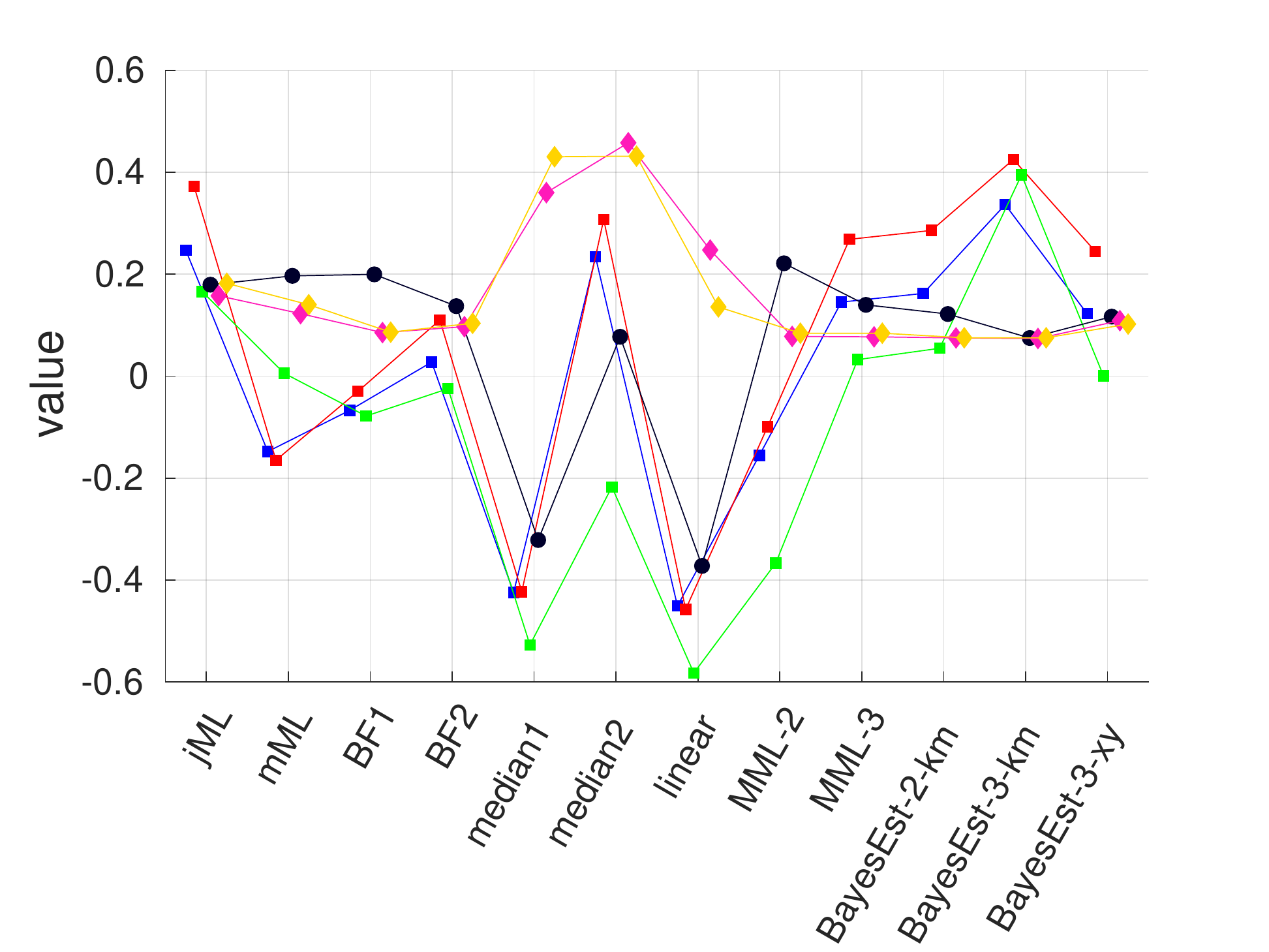} \\
  \ \\
  (c) predicted value for $N=2^4$ & (d) predicted value for $N=2^{13}$ \\
  \includegraphics[width=0.5\linewidth]{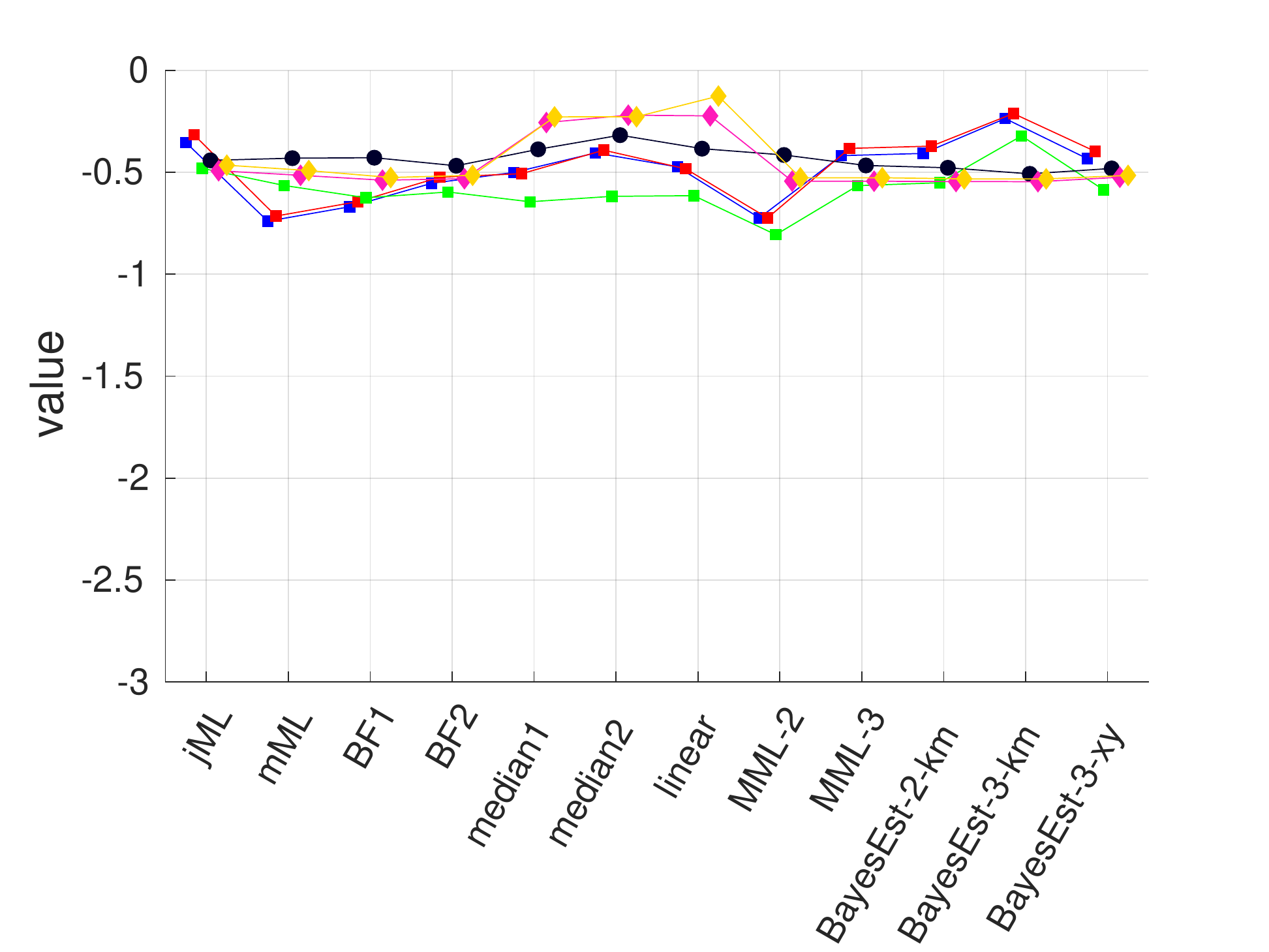} 
  & \includegraphics[width=0.5\linewidth]{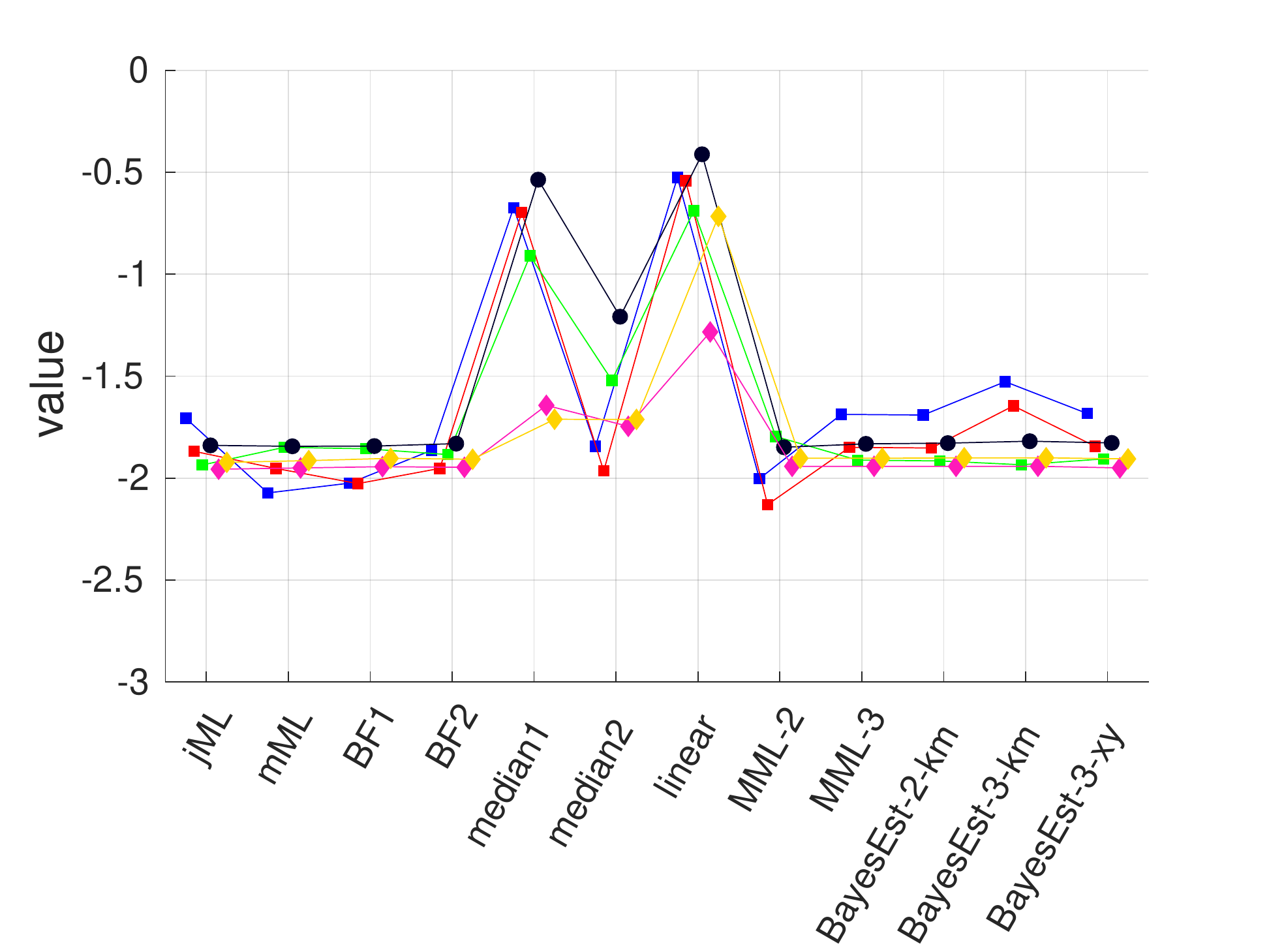} \\
  \ \\
  (e) standard deviation of residual noise & \\
  \includegraphics[width=0.5\linewidth]{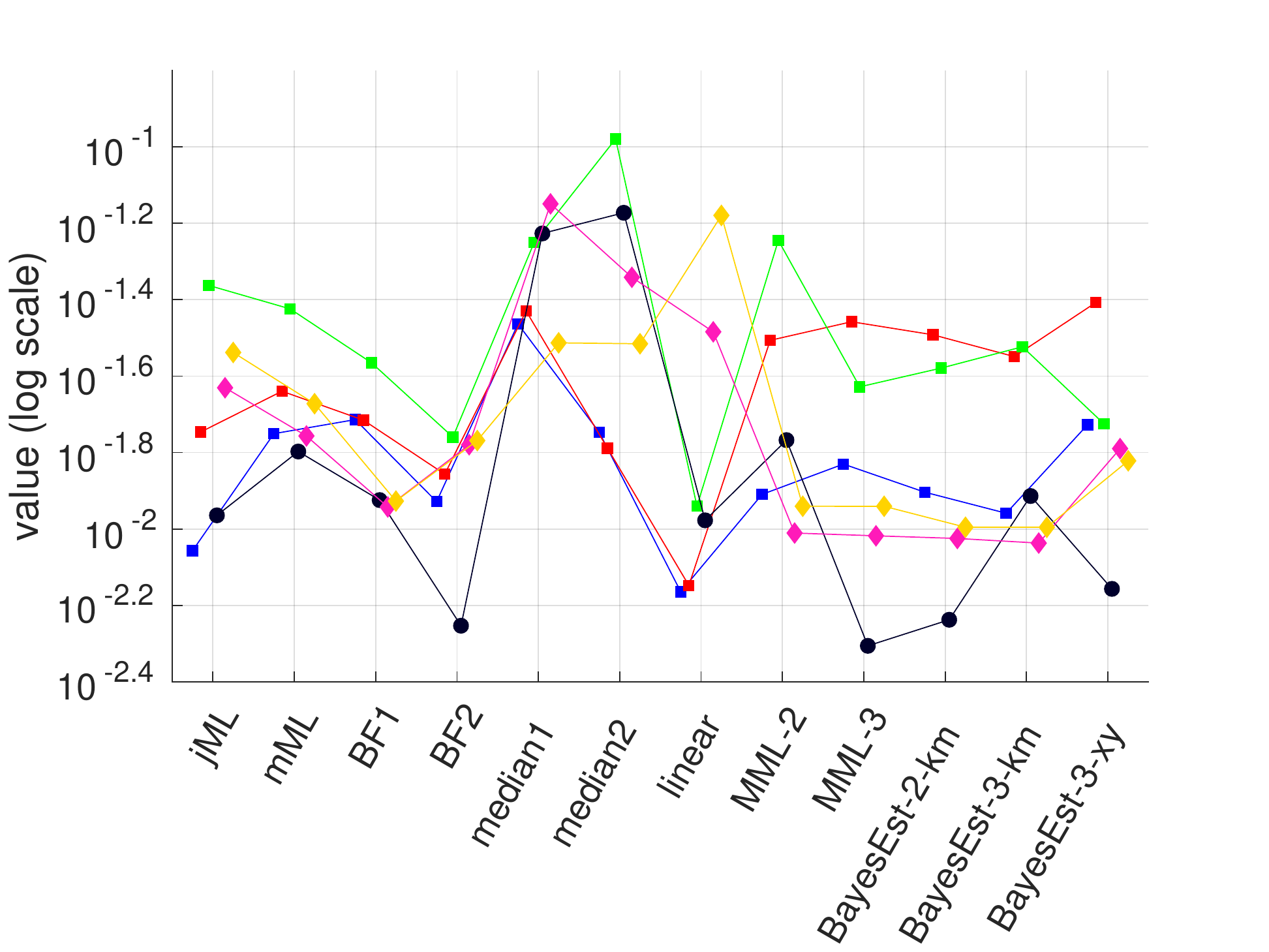} 
 \end{tabular}
 \caption{\textbf{Simulation study.} Result of linear regression analysis ($N \geq 2^4$) for either Equation~(\ref{eq:sim:reg:1}) (for $\tau \leq 1$) or Equation~(\ref{eq:sim:reg:2}) (for $\tau \geq 1$). (a) Estimated slope $\alpha$. (b) Estimated intercept $\beta$. (c) Predicted values of $\log_{10} ( \overline{\epsilon} )$ (for $\kappa \leq 1$) or $\log_{10} ( \overline{\epsilon'} )$ (for $\kappa \geq 1$) at $N = 2^4$. (d) Predicted values of $\log_{10} ( \overline{\epsilon} )$ (for $\kappa \leq 1$) or $\log_{10} ( \overline{\epsilon'} )$ (for $\kappa \geq 1$) at $N = 2^{13}$. (e) Standard deviation of residual noise.} \label{fig:sim:reg}
\end{figure}

To quantify these large-sample behaviors, we fitted to each curve a model of linear regression, either
\begin{equation} \label{eq:sim:reg:1}
 \log_{10} ( \overline{\epsilon} ) = \alpha \log_{10} ( N ) + \beta + \eta, \qquad N \geq 2^4, 
\end{equation}
for $\kappa \leq 1$, or
\begin{equation} \label{eq:sim:reg:2}
 \log_{10} ( \overline{\epsilon'} ) = \alpha \log_{10} ( N ) + \beta + \eta, \qquad N \geq 2^4,
\end{equation}
for $\kappa \geq 1$. The results are summarized in Figure~\ref{fig:sim:reg}, confirming generally homogeneous behaviors (at the exception of \texttt{median1}, \texttt{median2} and \texttt{linear}) that tend to depend more on the value of $\kappa$ than on the specifics of the estimation method.

\paragraph{Small sample behavior.}

Behaviors were much more variable for small sample sizes, with both a variability between methods and, for a given method, between values of $\kappa$. A majority of estimators (\texttt{jML}, \texttt{mML}, \texttt{BF1}, \texttt{median1}, \texttt{median2}, \texttt{MML-2}, \texttt{MML-3}) tended to have vey large average errors for very small sample sizes ($N \in \{ 2, 4 \}$). It was found that these extremely large values were due to a few very large errors. For instance, in the case of \texttt{MML-2}, $\tau = 1$ and $N = 2$, all but one of the 1\,000 repetitions had errors smaller than 1, the remaining error being equal to $3.72 \times 10^8$.
\par
\begin{figure}[!htbp]
 \centering
 \begin{tabular}{cc}
  (a) estimated decay amplitude $\gamma$ & (b) estimated decay time $\tau$ \\
  \includegraphics[width=0.5\linewidth]{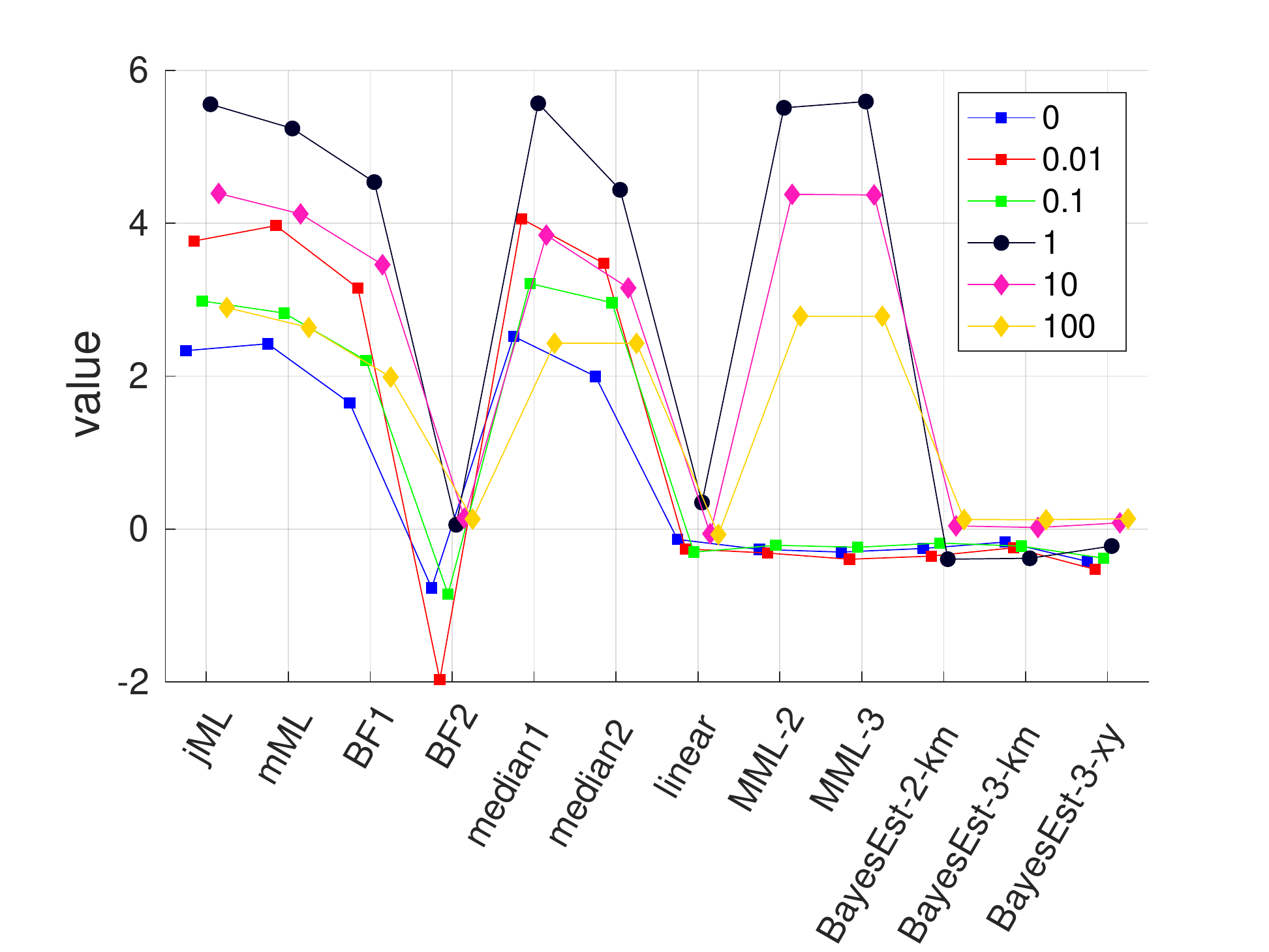}
  & \includegraphics[width=0.5\linewidth]{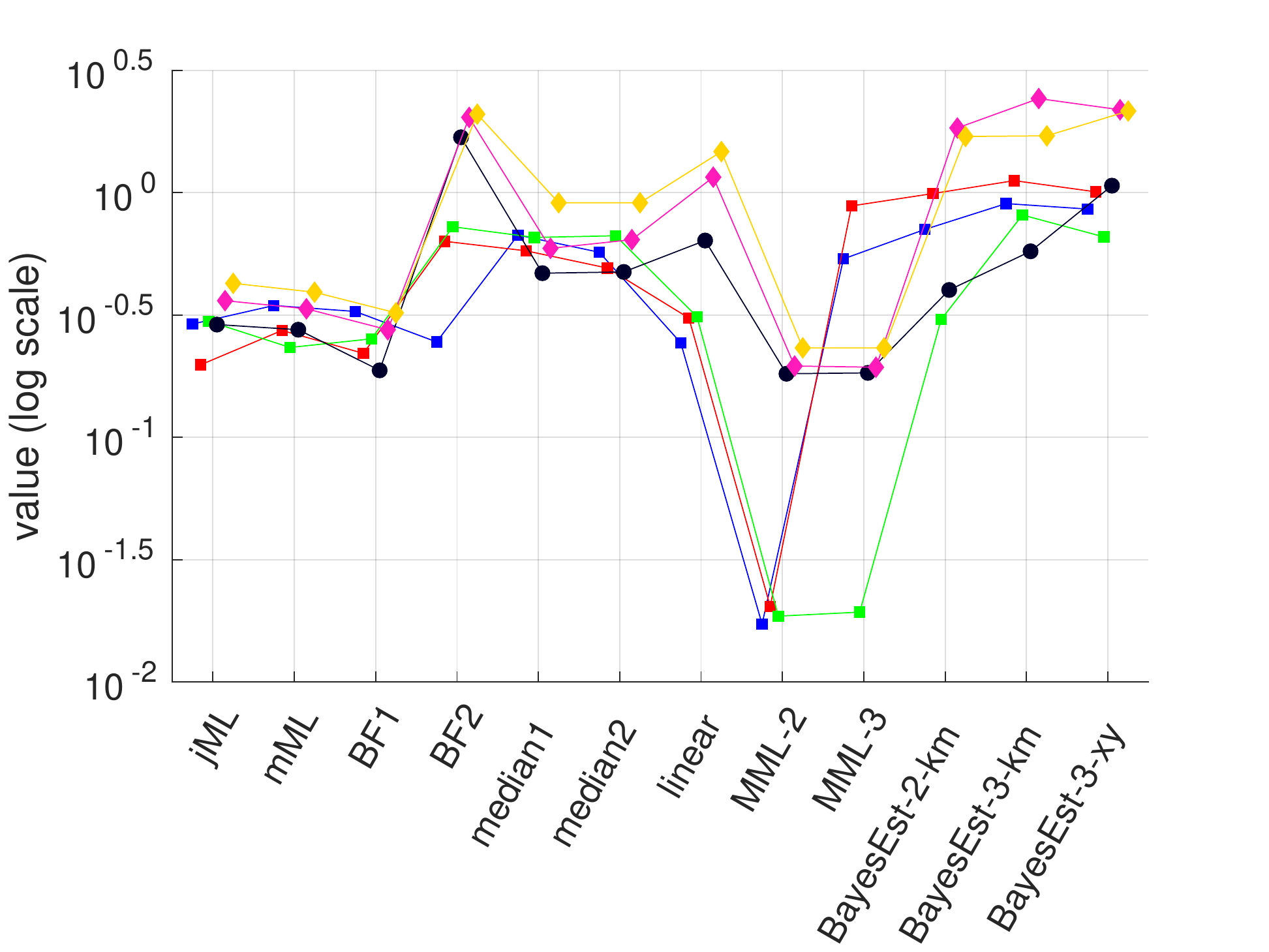} \\
  \ \\
  (c) standard deviation of residual noise & \\
  \includegraphics[width=0.5\linewidth]{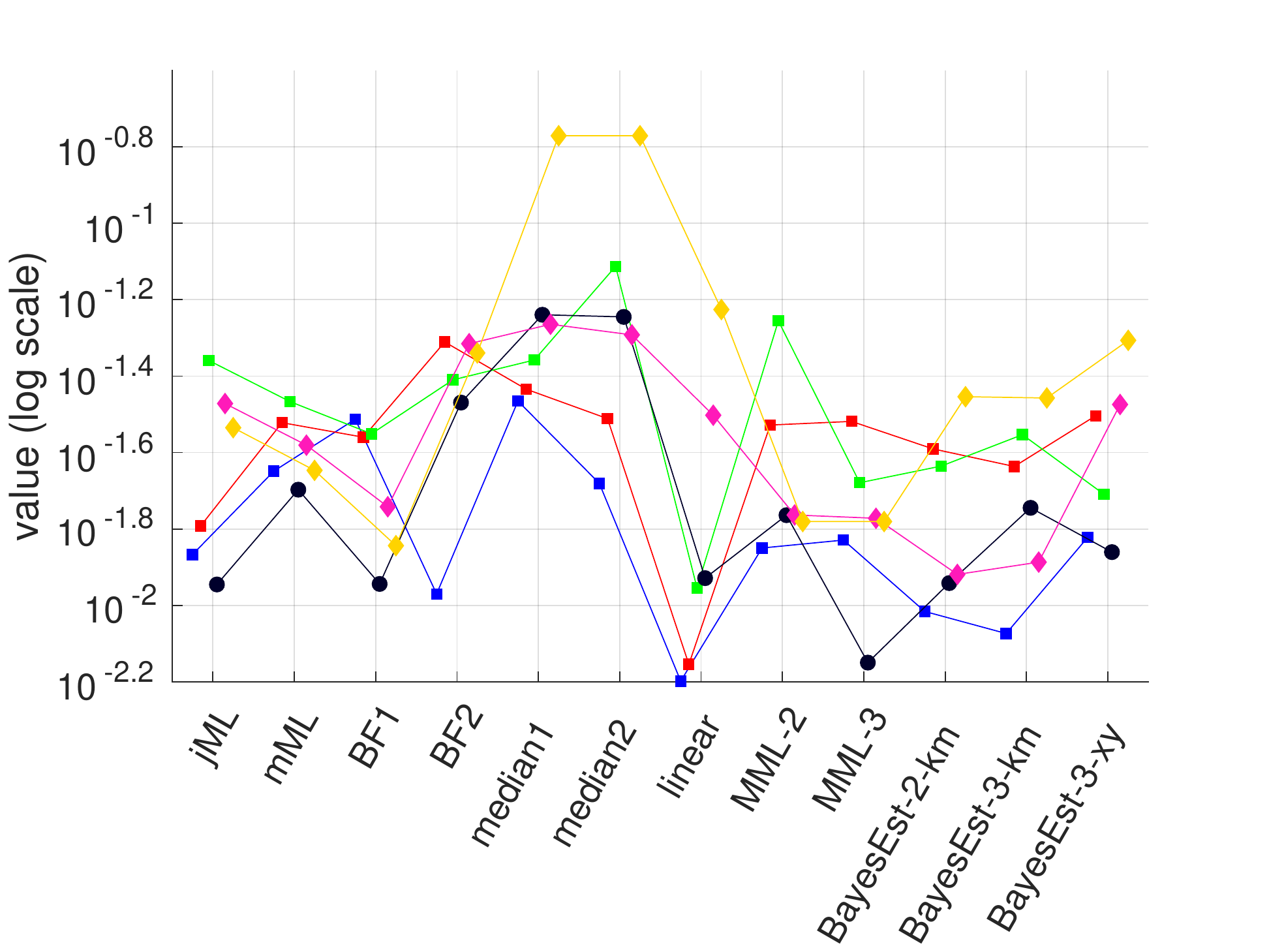} 
 \end{tabular}
 \caption{\textbf{Simulation study.} Result of nonlinear regression analysis for either Equation~(\ref{eq:sim:exp:1}) (for $\tau \leq 1$) or Equation~(\ref{eq:sim:exp:2}) (for $\tau \geq 1$). (a) Estimated decay amplitude $\gamma$. (b) Estimated decay time $\tau$. (c) Standard deviation of residual noise (bottom left).} \label{fig:sim:exp}
\end{figure}

To quantify this behavior, we modeled deviation from the linear trend as an exponential decay, either
\begin{equation} \label{eq:sim:exp:1}
 \log_{10} ( \overline{\epsilon} ) = \wh{\alpha} \log_{10} ( N ) + \wh{\beta} + \gamma \left( \frac{1}{10} \right) ^ { \frac{ \log_{10} ( N ) - \log_{10} ( 2 ) } { \tau } } + \eta
\end{equation}
for $\kappa \leq 1$ or
\begin{equation} \label{eq:sim:exp:2}
 \log_{10} ( \overline{\epsilon'} ) = \wh{\alpha} \log_{10} ( N ) + \wh{\beta} + \gamma \left( \frac{1}{10} \right) ^ { \frac{ \log_{10} ( N ) - \log_{10} ( 2 ) } { \tau } } + \eta
\end{equation}
for $\kappa \geq 1$. $\gamma$ is the maximum departure of the mean (relative) absolute error from the linear trend and corresponds to the increase in mean (relative) absolute error at $N = 2$. A positive value indicates a mean (relative) absolute error that is larger than the linear expectation (and therefore a decreased performance), while a negative value indicates a smaller mean (relative) absolute error (and therefore an increased performance). $\tau$ is such that the departure from the linear trend is divided by 10 for $\log_{10} ( N ) = \log_{10} ( 2 ) + \tau$. The larger the value of $\tau$, the slower the decay. The results are summarized in Figure~\ref{fig:sim:exp}.
\par
Regarding $\gamma$, we found several groups of estimators. A first group (\texttt{jML}, \texttt{mML}, \texttt{BF1}, \texttt{median1}, \texttt{median2}) had large positive values of $\gamma$ regardless of $\kappa$. A second group (\texttt{MML-1} and \texttt{MML-2}) had large positive values of $\gamma$ only for $\kappa \geq 1$, and small negative values for $\kappa < 1$. A third group (\texttt{linear}, \texttt{BayesEst-2-km}, \texttt{BayesEst-3-km}, \texttt{BayesEst-3-xy}) had low (positive and negative) values of $\gamma$ for all values of $\kappa$. A last estimator (\texttt{mML}) had small positive values of $\gamma$ for $\kappa \geq 1$ and negative, small yet larger values of $\gamma$ for $\kappa < 1$.
\par
Regarding $\tau$, all values were found to be in the range $[0.01, 3.2]$. All estimators with large values of $\gamma$ (corresponding to the first two groups mentioned earlier) had values in the range $[0.01, 1]$, indicating a very fast exponential decay --- for practical purposes, it mostly affected datasets of size $N = 2$.

\section{Discussion}

In the present manuscript, we assessed the behavior of 12 estimators of the concentration parameter of a von Mises distribution. We used synthetic datasets of size $N$ ranging from 2 to 8\,192 generated with a $\kappa$ ranging from 0 to 100. We provided detailed results as well as a more global view of the estimators' performance. We found that, for a given estimator, results were very similar across values of $\kappa$ if we considered the mean absolute error for $\kappa \leq 1$ and the mean relative absolute error for $\kappa \geq 1$.  We also found that, for a given $\kappa$, most estimators had similar behaviors for large datasets ($N \geq 2^4$), while the behaviors differed more strongly for small datasets.
\par
The fact that the mean absolute error for $\kappa \leq 1$ was similar to the mean relative absolute error for $\kappa \geq 1$ has two consequences. It first shows that, for $\kappa \geq 1$, error increases linearly as a function of $\tau$. Larger values of $\tau$ are therefore expected to lead to larger estimation errors. While it also means that smaller values of $\tau$ tend to yield smaller estimation errors, this behavior cannot be used to our advantage for $\kappa \to 0$, where it would mean that error vanishes as $\kappa \to 0$. By contrast, it is the mean absolute error that remains stable for $\kappa \leq 1$, i.e., error reaches a plateau when $\kappa$ decreases below $\kappa = 1$.
\par
For large datasets, the common behavior of most estimators was found to be that the mean absolute error roughly decreased linearly in log-log coordinates, with a slope around $-1/2$, corresponding to a decrease of mean (relative) absolute error in $1/\sqrt{N}$. This is in line with the general estimation theory, where errors are often found to decrease as $1/\sqrt{N}$.
\par
In the light of our results, we are able to give general practical guidelines regarding the use of the estimators tested here. First, we would not recommend the use of \texttt{median-1}, \texttt{median-2} and \texttt{linear} unless there is evidence that they might perform well in the specific context of interest. For large datasets ($N \geq 2^4$), all other estimators perform in a similar fashion and there is no obvious reason to recommend one or the other. In this context, \texttt{BF2}'s computational burden is a disadvantage with no compensation in terms of performance. By contrast, for small datasets, \texttt{BF2} was found to be the estimator with the best behavior, still at the price of a high computational cost as well as a consistent underestimation of the concentration parameter when it was small but different from 0. For instance, for $N = 2$, it estimated $\kappa$ as 0 regardless of its actual value (supplemental material, Table~5). The Bayesian estimators, which only performed a little worse than \texttt{BF2} but had computational costs similar to the other estimators, could prove valuable trade-offs.
\par
Importantly, all estimators considered here require a sample $\{ x_1, \dots, x_N \}$ of $N$ independent and identically distributed (i.i.d.) realization of a $\mathrm{vM} ( \mu, \kappa )$ distribution with $\mu$ and $\kappa$ unknown. Our simulation study faithfully respected this requirement. As a consequence, it did not explore the behavior of estimators with respect to the presence of outliers or model misspecification. Another consequence of the fact that we did not depart from the von Mises model is that we did not consider estimators assuming that one or several observations are identified as outliers (e.g., Winsorized estimate of \gcitep{Fisher_NI-1982}). Also, some of the estimators proposed in the literature depend on parameters (see, e.g., \gcitep{Lenth-1981, Kato_S-2016}). Since the behavior of these estimators strongly depends on the choice of the parameters, they were not incorporated in the analysis.
\par
As mentioned earlier, there were cases where some of the estimators were not defined. This happened for the linear estimator and $N = 2$, since this estimator requires at least 4 data points. The median-based estimator $\hat{\kappa}_{\med}$ also failed sometimes, in particular for small values of $\kappa$, as Equation~(\ref{eq:def:median2}) had no solution (the integral was smaller than 0.5 for all values of $\kappa \geq 0$). In such cases, we could have decided to set the estimator to 0. However, we believe that such problem is symptomatic of a deeper problem whose investigation goes beyond the scope of the present manuscript. So we just took the estimator as is.
\par
As a final note, the routines calculating the various estimators were written by us and were not optimized. As such, the computation times presented here should only serve as a rough indicator of the time required to apply each of them. In particular, it is possible that optimal coding of \texttt{BF2} with a compiled language could greatly improve its computation time.

\bibliographystyle{chicago} 
\bibliography{nonabrev,anglais,mabiblio}

\end{document}